\documentclass[conference,a4paper]{IEEEtran}
\usepackage[singlespacing]{setspace} 
\IEEEoverridecommandlockouts
\usepackage{cite}
\usepackage{amsmath,amssymb,amsfonts}
\usepackage{algorithm}
\usepackage{algpseudocode}
\usepackage{graphicx}
\usepackage{textcomp}
\usepackage{blindtext}
\usepackage{subfigure}
\usepackage{float} 
\usepackage{subfigure}
\usepackage{xcolor}
\usepackage{diagbox}
\usepackage{slashbox}
\usepackage{wrapfig}
\usepackage{makecell}
\usepackage[left=1.7cm,right=1.7cm,top=1.9cm]{geometry}
\def\BibTeX{{\rm B\kern-.05em{\sc i\kern-.025em b}\kern-.08em T\kern-.1667em\lower.7ex\hbox{E}\kern-.125emX}}
\makeatletter

\begin{document}
	\title{A Short Overview of 6G V2X Communication Standards}
	\IEEEpeerreviewmaketitle
	\author{\IEEEauthorblockN{Donglin Wang\IEEEauthorrefmark{1}, Yann Nana Nganso\IEEEauthorrefmark{1}, and Hans D. Schotten\IEEEauthorrefmark{1}}
		\IEEEauthorblockA{\textit{\IEEEauthorrefmark{1}Rhineland-Palatinate Technical University of Kaiserslautern-Landau, Germany} \\
		$\{$dwang,schotten$\}$@eit.uni-kl.de \\
          $\{$nganso$\}$@rhrk.uni-kl.de }
  }
	\maketitle
\begin{abstract}
We are on the verge of a new age of linked autonomous cars with unheard-of user experiences, dramatically improved air quality and road safety, extremely varied transportation settings, and a plethora of cutting-edge apps. A substantially improved Vehicle-to-Everything (V2X) communication network that can simultaneously support massive hyper-fast, ultra-reliable, and low-latency information exchange is necessary to achieve this ambitious goal. These needs of the upcoming V2X are expected to be satisfied by the Sixth Generation (6G) communication system. In this article, we start by introducing the history of V2X communications by giving details on the current, developing, and future developments.  We compare the applications of communication technologies such as Wi-Fi, LTE, 5G, and 6G. we focus on the new technologies for 6G V2X which are brain-vehicle interface, blocked-based V2X, and Machine Learning (ML). To achieve this, we provide a summary of the most recent ML developments in 6G vehicle networks. we discuss the security challenges of 6G V2X. We address the strengths, open challenges, development, and improving areas of further study in this field.
\end{abstract}


\section{Introduction}
Vehicle-to-Everything (V2X) communications have the potential to drastically reduce the number of vehicle collisions and as a result the number of deaths \cite{naik2019ieee}. In recent years, V2X communications have attracted significant research interest from both academia and industry \cite{noor20226g}. V2X communications consist of Vehicle-to-Vehicle (V2V), Vehicle-to-Pedestrian (V2P), Vehicle-to-Infrastructure (V2I), and Vehicle-to-Network (V2N) communications are examples of V2X communications that increase road safety, traffic efficiency, and the availability of infotainment services.\cite{chen2017vehicle}.  

The Federal Communications Commission (FCC) of the United States \cite{garcia2021tutorial } allocated 75 MHz of spectrum in the 5.9 GHz band for Intelligent Transportation Services (ITS) in 1999. This development enhanced significant research activity around the technology of V2X communications. The results gave the first set of radio channels for V2X completed in 2010 \cite{garcia2021tutorial}. These standards, known as Dedicated Short Range Communications (DSRC) are based on IEEE 802.11p technology. Following the creation of radio standards, higher-layer standards, message formats, protocols, and applications emerged \cite{garcia2021tutorial}. The 3rd Generation Partnership Project (3GPP) enables through Release 12 (Rel.12) the availability of Device-to-Device (D2D) communications in Long Term Evolution (LTE), therefore, allowing competitive broadband communication technology for public safety networks \cite{lin2014overview}. According to \cite{chen2017vehicle} the achievement of better system performance of V2X services and LTE-based V2V is due to Release 14 (Rel.14). Rel.14 introduces LTE communication technology into vehicular networks (denoted LTE-V2X). It functions in two modes: network-based mode, which uses the LTE-Uu interface as the interface between a vehicle and a network infrastructure, and direct mode, which is based on D2D communications defined in 3GPP Release 13 (Rel.13), allowing devices to directly communicate (eg V2I, V2P) through the PC5 interface, known as the LTE sidelink considering the network infrastructure is not involved \cite{bagheri20215g}. In 2018, the V2X 5G New Radio (5G NR) technology was launched with the intention of enabling functions like vehicle platooning, advanced driver assistance, and remote driving. Furthermore, enhancements were made to the PC5 interface to increase its reliability, reduce latency, and boost data rates through methods such as carrier aggregation and advanced modulation techniques like 256QAM. Release 16 (Rel.16) is one of the few versions created by 3GPP on a cellular V2X based on 5G. It should be mentioned that Rel.16 seeks to deliver Ultra Reliable Low Latency Communication (URLLC) and better throughput while also enhancing and optimizing Rel.15 capabilities. 

3GPP is actively developing Release 17 (Rel.17), which will include architectural improvements to accommodate sophisticated V2X devices. URLLC is only allowed to operate in the licensed spectrum in Rel.15 and Rel.16. Due to the high demand for data transmission in 5G unlicensed spectrum becomes helpful in the licensed spectrum which suffers from low-cost bandwidth. The characteristics of the unlicensed spectrum do not take into consideration the URLLC in Rel.15 and Rel.16. This is why Rel.17 was designed to unify the characteristics of the unlicensed spectrum with URLLC, allowing URLLC to function at the expected latency and dependability. \cite{le2020overview}. In Fig. 1 we can see the unlicensed evolution of 3GPP \cite{baena2020kqi}.

\begin{figure*}[htbp]
    \centering
    \includegraphics[width=\linewidth]{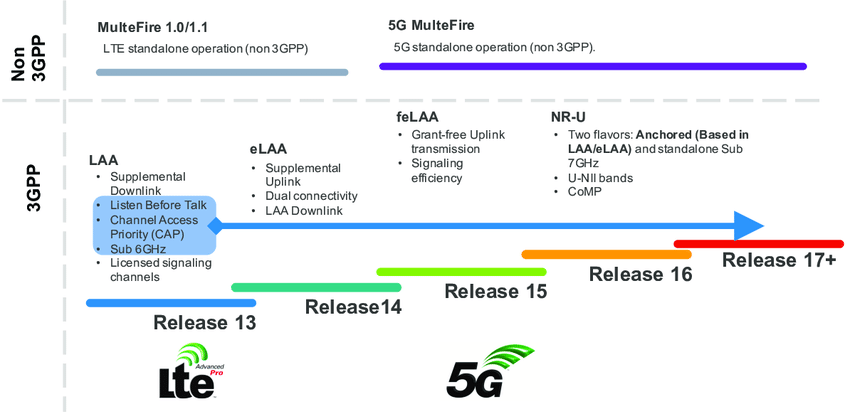}
    \caption{3GPP unlicensed evolution}
\end{figure*}

3GPP intends to improve the coverage of physical random access channels and approaches to boost User Equipment (UE) power in Release 18 (Rel.18). In addition, Rel.18 will introduce mechanisms to procedures for layer 1/layer2 based intercell mobility and will enable multi-SIM UE to indicate its preferred capability restriction to network A when the UE needs to communicate with network B. Furthermore, Rel.18 will study low-power wake-up receiver architectures; this procedure will support the wake-up receiver and the signal. It will also support 5G NR for devices onboard aerial aircrafts, such as Unmanned Aerial Vehicles (UAVs). Fig. 2 shows the evolution of 3GPP Releases \cite{lin2022overview}. The main contributions of this paper include the following:
\begin{itemize}
    \item Present the current and future development in V2X Communication.
    \item Examine and contrast the main performance characteristics of current vehicle technology with the anticipated needs for vehicular communication.
    \item Describe the developments in V2X communication now and in the future.
    \item Evaluate the security issues in V2X vehicular systems
\end{itemize}

This study is an attempt to suggest a description of the 6G V2X Communication standards. The paper is organized as follows. Section II will include the most current communication innovations connected to V2X. The use of 6G-V2x will be explained in Section III. In Section IV, we will present the new technologies for 6G V2X. Section V explains the security difficulties in V2X communication.

\section{Vehicle communication technologies}
The communication technologies examined in this study are described in this section.

\subsection{Dedicated Short-Range Communication}
Vehicles and infrastructure are able to establish secure, swift, and direct communication using DSRC, which is a wireless networking technology based on a variation of WiFi. DSRC operates within the 5.85 to 5.925 GHz frequency band, a dedicated spectrum reserved for vehicular communication with the Intelligent Transportation System (ITS). \cite{bey2019evaluation}.

\subsection{Long Term Evolution in 4G}

LTE is a cellular communication network that was developed by the Third Generation Partnership Project (3GPP) to address V2X communication concerns \cite{bey2019evaluation}. Cellular V2X (C-V2X) was launched by 3GPP in 2016 with the combination of LTE technology with vehicular communication \cite{filippi2017ieee802}.

\begin{table*}[htbp]
    \caption{Vehicular Communication Technologies and Key Performance Indicators}
    \label{tab1}
    \centering
    \begin{center}
    \begin{tabular}{|c|c|c|c|c|}
     \hline
     {\textbf{Application requirements}} & {\textbf{6G}} & {\textbf{5G}} & {\textbf{4G LTE}} & {\textbf{DSRC}}\\
     \hline
     Frequency & up to 1 THz & 3-300 GHz & 1.88-1.9 GHz & 5.85-5.925 GHz \\
     \hline
     Mobility & Up to 1000 km/h & Up to 500 km/h & Up to 140km/h & Up to 40 km/h\\
     \hline
      Satellite Integration & yes & No & No & No \\
     \hline
      Extreme Data Rate &  100 Gbps & 1Gbps &  100 Mbits/s & 27 Mb/s \\
     \hline
     Ultra-micro latency & $<$1ms & 10ms & $<$100ms & 150ms \\
     \hline
     Artificial Intelligence & Yes & No & No & No \\
     \hline
     Live Traffic & Yes & No & No & No \\
     \hline
     Big Data handling & Yes & No & No & No \\
     \hline
     Real-time hazard updates & Yes & No & No & No \\
     \hline
     Sharing HD 3 Dimension maps & Yes & No & No & No \\
     \hline
        
    \end{tabular}
    \end{center}
\end{table*}

\subsection{5G}
3GPP began the 5th generation mobile network, which focuses on the digitization of industry and servicing customers. According to \cite{andrews2014will} 5G enhances c-V2X communication with a capacity increase of 1000-fold, a data throughput of 1 Gbps, ample bandwidth in densely populated areas, improved responsiveness, decreased latency, and heightened reliability.

\subsection{6G}
Due to the inclusion of smart applications and modern technologies, vehicular communication equipment is becoming more complex. When it comes to speed, 6G surpasses 5G by a factor of 100. Additionally, 6G is considered the most intelligent and efficient cellular network compared to its predecessors. Consequently, 6G has been designed as the initial generation of cellular networks to rectify the limitations of the 5G network.

\section{Why 6G-V2X?}


For around two years, 5G in the frequency range 1 (FR1, or sub-6 GHz bands) has been effectively implemented all over the world. However, because mmWave access is limited to a smaller number of places, primarily in densely populated areas, the Return on Investment (ROI) of millimeter-wave (mmWave) 5G deployment in the frequency range 2 (FR2: 24-40 GHz) has just recently begun. mmWave radio access is a very promising opportunity for operators and manufacturers in the telecommunications sector to offer users new, enticing services, not just for high-performance use cases but as a technology enabler of smart cities supported by a knowledgeable ecosystem with robots, tactile internet, Virtual and Augmented Reality (VAR), and pervasive ITS. Artificial Intelligence (AI) at all levels is bracing these and other industries, along with the expanding availability of reliable, safe, and enormously exploitable data sources \cite{mizmizi20216g}.
Given this perspective, it is possible that 5G NR-based V2X networks will not be able to support all of the necessary specifications and use cases. Furthermore, while the ideas behind ITS have been extensively examined for many years, legacy V2X communication technologies can only offer a certain amount of intelligence. As a result, a significant paradigm change is required away from traditional communication networks and toward more adaptive and diversified network strategies. This transformation is expected to begin with the recently announced 6G wireless communication network, which would incorporate a variety of non-terrestrial communication networks, including satellite and UAV communication networks. As displayed in Table \ref{tab1}, the 6G-based V2X technology demonstrates a superior capability in addressing a wide array of application requirements in comparison to its traditional V2X communication counterparts. This advancement in 6G not only brings faster speeds and lower latencies but also enhances the overall efficiency and reliability of V2X communications, paving the way for more innovative applications and improved user experiences in the realm of vehicular communication. 

With billions of connected communication devices, this will allow for truly intelligent and ubiquitous V2X systems with greatly improved reliability and security, extremely high data rates (such as Tbps), massive and hyper-fast wireless access, and much smarter, longer, and greener (energy-efficient) Three-Dimensional (3D) communication coverage. 

Due to the very diverse composition of the network, variable communication situations, and stringent service requirements, future V2X networks will require unique approaches to facilitate adaptive learning and intelligent decision-making. The combination of 6G with ML is expected to bring about several new characteristics, including improved context awareness, self-aggregation, adaptive coordination, self-configuration, and the full potential of radio signals by evolving to intelligent and autonomous radios \cite{noor20226g}.

\begin{figure*}[htbp]
   \centering
   \includegraphics[width=\linewidth]{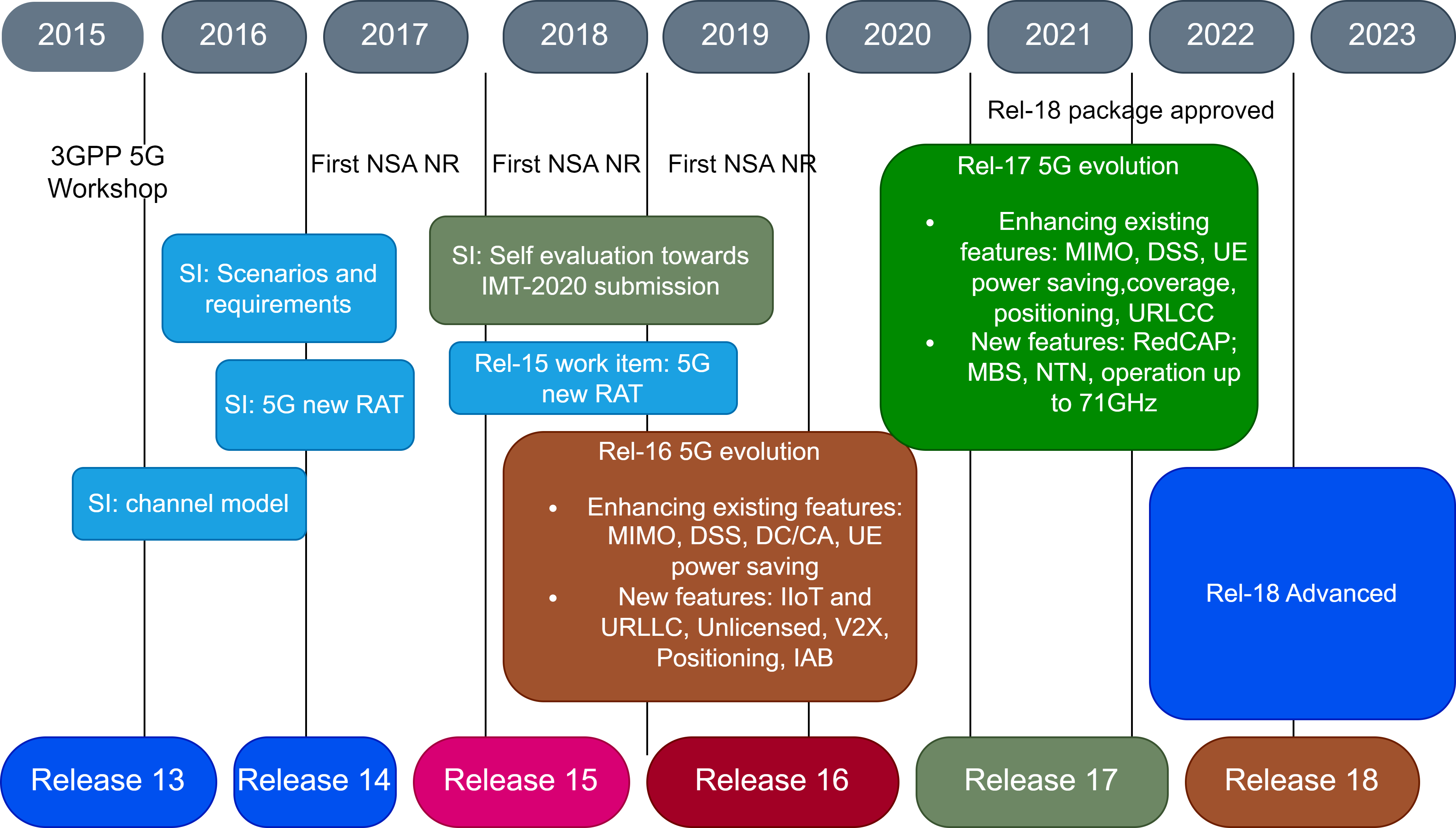}
  \caption{3GPP's 5G evolution roadmap}
 \end{figure*}

\section{New technology for 6G-V2X}
This section introduces some of the cutting-edge technologies that could be used in 6G-V2X.

\subsection{Brain-Vehicle Interfacing}
Instead of a physical link between the driver and the vehicle, a Brain-Controlled Vehicle (BCV) is driven by the driver's thoughts. BCV may considerably boost freedom for people with disabilities by providing them with an additional interface for controlling automobiles. The brain-vehicle interface has the potential to improve manual driving by anticipating a driver's actions and sensing discomfort. While the ultimate aim is fully automated vehicles, managing the unpredictability and intricacies of autonomous driving will necessitate human adaptability. A brain-vehicle interface is expected to mitigate the challenges of autonomous driving in challenging and uncertain situations, like rural and less organized areas, by maintaining human involvement. While services connected to brain-machine interactions would demand concurrent ultrahigh dependability, ultralow latency, ultrahigh data rate connectivity, and ultrahigh-speed computing, current wireless communications (e.g., 5G) and compute technologies are unable to actualize BCV. For instance, a rough estimate of the full brain recording requirement puts it at roughly 100 Gb/s, which present wireless technology cannot handle. However, to assist learning and adapt to the behavior of human drivers, 6G-V2X must use a fully phased brain-vehicle interface and ML approaches.

\subsection{UAV/Satellite-Assisted V2X}
Due to their extensive aerial coverage, UAVs have the potential to act as airborne wireless access points within the 6G-V2X network. These unmanned aerial vehicles can offer various services to mobile customers, such as relaying data, storage, and computational capabilities. UAVs can work together with base stations and other components of the immobile network infrastructure, especially in highly congested vehicle environments, to manage the wireless network and improve the user experience. Unmanned Aerial Vehicles (UAVs) offer numerous distinctive V2X applications as airborne agents, benefiting from their nearly unrestricted three-dimensional mobility. These applications include: 1) Offering early reports of road accidents prior to the arrival of rescue teams. 2) Surveillance of traffic violations to support law enforcement agencies; and 3) Disseminating warnings regarding road hazards in regions lacking pre-installed Roadside Units (RSUs). Another possible airborne communication vehicle for 6G-V2X communications is to use satellites. Satellites are currently utilized in the V2X specifications for localization. It is important to note that satellite transmission data rates have increased significantly recently. Multibeam satellites, for example, have been widely employed in satellite communication networks due to their ability to boost wireless transmission throughput. As a result, 6G-V2X may use satellite communication as a method to help contact between a car and a distant data center in a situation where there is no terrestrial service. Similar to UAV-based V2X communication, a satellite can also perform network and computing tasks \cite{noor20226g}.

\subsection{Integrated Control and Communication}
A key component of 6G will be integrated communication and management, which may also contribute to the advancement of sophisticated and autonomous V2X services. One use of integrated communication and management is vehicle platooning, in which a group of automobiles travels closely together in a car-coordinated motion without any mechanical connection. The capacity of the road is increased, fuel economy improves and pleasant road trips are three of the main advantages of car platooning. To synchronize their acceleration and slowdown, each vehicle in the platoon needs to know the relative distance and velocity of its nearby vehicles \cite{noor20226g}.

\subsection{Blocked-Based V2X}
The wide adoption of V2X networks is highly dependent on authentication and message propagation for large-scale vehicle traffic. This concept sets new restrictions on how resources may be distributed in V2X networks. To combat potential malicious assaults or jamming, mission-critical communications, for instance, should have ultra-resilient security, but multimedia data services may simply need lightweight protection owing to the volume of data. Different frame structures, routing,  power, and spectrum allocation techniques result from these two different kinds of security needs. 6G-V2X might use a blockchain system, which is considered a disruptive technology for safe multiparty decentralized transactions. The utilization of blockchain technology presents a range of enhanced security and privacy services, distinct from traditional security solutions, as it eliminates the need for intermediaries. Blockchain's inherent distributed ledger technology enables decentralized security management, offloading tasks to mobile cloud/edge/fog computing, and caching content within 6G-V2X communication. In the context of 6G-V2X, a blockchain-based security solution, such as a smart contract or consensus mechanism, is expected to not only verify message authenticity but also safeguard the sender's privacy. Blockchains also play a pivotal role in managing unlicensed spectrum, enabling multiple users to share the same spectrum. 6G-V2X may adopt a blockchain-based approach to spectrum sharing, potentially delivering a more secure, intelligent, cost-effective, and highly efficient decentralized spectrum-sharing solution. Although there have been various attempts to implement a blockchain-based communication network, the dynamic nature of the V2X communication situation and the need for real-time data processing preclude a straightforward adoption of existing blockchain technology. The technology itself suffers from excessive latency, despite blockchain's enormous promise to provide improved security and network management. Consequently, the development of new blockchain algorithms with ultra-low latency is imperative before their integration into 6G-V2X can be realized. Some significant outstanding issues with existing blockchain technology that require in-depth research include its limited throughput and scalability.

\subsection{Machine Learning}
Recent developments in ML studies have allowed the development of new technologies such as self-driving cars and speech assistants due to the availability of large data sets, storage and computing capacity. Given this context, ML has become more and more crucial to the highly independent and clever functioning of tomorrow's 6G vehicular networks. Traditional wireless communication system design is heavily dependent on model-based methodologies, in which different communication system building elements are carefully built based on analysis of measurement data. Although these model-based methods have proven effective in designing conventional communication systems, there may be some 6G-V2X situations where precise modeling (such as an accurate interference model and channel estimation) is unlikely.In the previously mentioned scenarios, where traditional communication system design might encounter discrepancies in model compatibility, ML proves to be a potent instrument due to its ability to discern features and detect relationships, even those concealed deeply, between input and output data. Furthermore, the data-driven essence of machine learning can facilitate predictions and projections regarding user behavior, network traffic, application needs, security threats, and channel dynamics. This, in turn, leads to improved resource allocation and enhanced network performance. Recent advancements in machine learning (ML) methods are significantly contributing to the progress of autonomous vehicles, while also playing a crucial role in enhancing the overall driving experience and road safety. Take, for instance, the wealth of data streams originating from sources like cameras, Light Detection and Ranging (LiDAR) sensors, GPS units, and various sensors. These data streams can be efficiently processed, allowing for the application of modular perception, planning, action, or end-to-end learning techniques in automated driving, enabling data-informed intelligent decision-making.
In the realm of ML-based vision, there's ongoing exploration of multimodal reasoning. This involves the integration of camera frames and LiDAR scans to enhance object detection, a vital element in automated driving safety.Our focus in this paper is on the network perspective, highlighting the influence of ML within 6G-V2X networks. However, it's worth noting that numerous ML-driven applications for intelligent driving are anticipated in the future. We also introduce one of the most cutting-edge ML techniques, federated learning \cite{noor20226g}.
\subsubsection{ML for security management} 
The merge of various forms of communication and the strict data delivery requirements for 6G-V2X will worsen security problems.The inherent broadcast nature of vehicular communication exposes it to potential security vulnerabilities and malicious attacks, although 6G-V2X seeks to provide smooth access between infrastructure sites and cars. A vehicular network could be the subject of numerous malicious attacks, including approved and sanctioned attacks, data forgeries, and distributions \cite{sharma2020security}. A new user identification and verification method must be developed to maintain safe and legal access to data, services, and systems because private user information, such as user identity or trajectory, is shared over wireless links in a V2X system \cite{machardy2018v2x}. In \cite{grover2011machine} and \cite{scalabrin2017bayesian}, supervised learning with categorization capacity is suggested as a useful tool to detect abnormal driving behavior in cars. It is stated that because training and detection processes rely on previously labeled data, supervised learning may not be able to identify new or undiscovered assaults. Unsupervised learning, which can aggregate data without the need for labeled information, is being considered for real-time recognition in \cite{maglaras2015novel} and \cite{sequeira2002admit}. For vehicle networks, intruder detection using K-means clustering is specifically suggested in \cite{maglaras2015novel}. Anomaly detection using unsupervised learning is researched in \cite{sequeira2002admit} to combat assaults that can occur spontaneously in real-time. Nonetheless, these approaches exclusively focus on either misuse detection or anomaly detection, which may not effectively address real-world scenarios where both known and undiscovered attacks could occur simultaneously. Additionally, to reduce transmission costs, reactive detection is taken into account primarily in current detection methods.Nonetheless, it is anticipated that proactive exploration-based security methods will be beneficial for elevating the security standards within a 6G-V2X network, particularly in scenarios where communication resources are relatively abundant \cite{tang2019future}. For example, \cite{al2017proactive} adopts a proactive anomaly detection strategy for linked vehicles to avoid cyber threats. Typically, cryptographic techniques are used in the higher layers of the protocol stack to handle security problems in wireless transmission. However, in diverse and dynamic V2X networks where cars may haphazardly enter or exit the network at any moment, the administration and exchange of private keys will be difficult \cite{elhalawany2019physical}. In this sense, Physical Layer Security (PLS) solutions can be used to complement conventional encryption methods \cite{liu2016physical}. While Physical Layer Security (PLS) techniques leverage the physical attributes and stochastic behavior of wireless channels to mitigate eavesdropping, they can still be influenced by the accuracy of channel modeling. In a V2X context marked by high mobility and frequent channel variations, the use of machine learning becomes advantageous for accurate channel prediction and monitoring. This, in turn, has the potential to enhance the effectiveness of PLS-based methods. Furthermore, various security levels are anticipated according to the situations and services. Consider, for instance, two cars that track one another either at a busy junction or on a deserted road. Due to the mobility of cars, the latter has a greater number of variables that can influence decision-making, leading to strict security requirements \cite{furqan2019intelligent}. In the latter scenario, ML may be used to dynamically determine the necessary security degree and the best PLS answer. As demonstrated in \cite{ferdowsi2019cyber}, ML can also play a role in enhancing the management and communication systems designed to safeguard against data injection attacks within specific vehicle platoons. When ML is used to improve security, the end-to-end network efficiency of the ML-based system must be validated. As stated above, ML can be applied to functional components at various network levels. To ensure that all interactions are secure, ML usage should be coordinated across the network \cite{ylianttila20206g}.

\subsubsection{Federated Learning for 6G-V2X}
Training ML models, which can be used at base sites or in vehicles, is a key problem for effective use of ML. Training a large ML model in distant clouds is an apparent answer, but it may take some time. One issue is that a sluggish reaction to external changes could result from rapidly changing vehicle network and communication conditions, which would affect performance. Furthermore, given that the majority of training data is generated at network edges, such as base stations and mobile vehicles, the expense and latency associated with transmitting this data to a remote cloud can be substantial. In view of this, conducting local training of machine learning models within 6G-V2X networks emerges as a more favorable choice. To increase the accuracy and generalization of the performance of ML models, joint training samples are a possible solution. This is because each base station or car may only be able to store a limited number of training samples.While base stations and vehicles may have concerns about compromising their privacy by sharing training data, privacy remains a significant concern in collaborative training. Federated learning, a relatively recent approach designed to address privacy and transmission overhead issues associated with ML model development, has garnered considerable research attention as a means to enhance wireless networks. \cite{park2019wireless} \cite{chen2020joint} \cite{yang2019federated} \cite{niknam2020federated}. There are numerous technological challenges to address when implementing federated learning applications, which are regarded as a promising machine learning approach for enhancing the efficiency of 6G-V2X networks. Supervised learning is taken into account primarily in the study that has already been conducted on shared learning in wireless networks. As the use of Reinforcement Learning (RL) models increases, a flexible federated RL system is required that can accommodate a variety of 6G-V2X use cases.  Furthermore, initiating federated RL (reinforcement learning) from a blank slate is often infeasible since many V2X applications are mission-critical, and such an approach could lead to an unstable initial phase during the learning process. The limited connection between vehicles presents another difficulty for federated learning involving cars. Vehicles may not be in contact with base units or other federated learning-related vehicles. Therefore, while they are stopped, the cars might need to participate in federated learning. Lastly, a more comprehensive study is needed to determine how the wireless route affects the effectiveness of federated learning. Wireless delays and errors can affect the precision of federated learning, as demonstrated in \cite{chen2020joint}. The high-speed mobility of vehicles and the dynamic nature of channels within a mobile V2X network could exacerbate this effect. Further research is essential to explore the integrated development of wireless and learning processes for V2X applications.

\section{V2X communication security}
When it comes to communication across public or private networks, one of the primary issues is security. When transmitting, receiving, and processing sensitive data over networks, network systems are susceptible to a range of threats, including denial of service attacks, eavesdropping, and challenges related to digital identity verification. V2X technology enables vehicles to autonomously detect and respond to rapidly changing objects as symbols. If the information exchanged within the vehicular network is accessed and modified without authorization, it could lead to catastrophic consequences for the vehicle's decision-making process. As an illustration, within an autonomous vehicle, remote control over components like the steering wheel and braking system is possible. If an unauthorized party gains control over these functions, it could lead to potential harm to both property and individuals. Furthermore, one of the most well-known ideas today that makes it easier to manage city traffic is the Intelligent Transport System. Among the most important traffic management practices are the enhancement of traffic flow through analysis of daily traffic patterns, traffic prioritization, and upkeep of intelligent traffic signals. The security of the community and the Intelligent Transportation System (ITS) is jeopardized when unauthorized individuals gain access to the data, and the entire network could be disrupted if any data is compromised or tampered with.

\section{Conclusion}
In this article, we identified several enabling technologies for 6G V2X communications. Our main focus was to understand how 6G-V2X functions and its applications. 6G technology excels in meeting the requirements for V2X communication. Nevertheless, security remains a significant challenge in V2X communication and must be effectively addressed to drive further advancements in 6G technology.d We hope that this article with provide specific details in 6G-based next-generation V2X for other scientists and engineers which will also encourage more research and solutions on this topic. 

\section{acknowledgement}
This work has been supported by the Federal Ministry of Education and Research of the Federal Republic of Germany (BMBF) as part of the Open6GHub project with funding number 16KISK004. The authors would like to appreciate the contributions of all Open6GHub partners. The authors alone are responsible for the content of the paper which does not necessarily represent the project.

\bibliographystyle{IEEEtran}
\bibliography{references}

\end{document}